\documentclass[aps,pre,twocolumn,superscriptaddress]{revtex4-2} 

\usepackage{graphicx} 
\usepackage{dcolumn}  
\usepackage{float}    
\usepackage{physics}
\usepackage{bm}       

\usepackage{amsmath}
\usepackage{amssymb}
\usepackage{amsthm}

\usepackage{booktabs} 
\usepackage{tabularx}

\makeatletter
\def\switch@array{}
\makeatother
\begin{document}

\preprint{APS/123-QED}

\title{Infectious Disease Induces Emergent Oscillations, Extinction and Changes in Community Persistence in a Food Chain}

\author{Hooman Saveh}
\affiliation{Department of Physics, Sharif University of Technology, Tehran, Iran}

\author{Fakhteh Ghanbarnejad}
\email{fakhteh.ghanbarnejad@gmail.com}
\affiliation{School of Technology and Architecture, SRH University of Applied Sciences Heidelberg, Campus Leipzig, Prager Str. 40, 04317 Leipzig, Germany}
\affiliation{Zuse Institute Berlin, Berlin, Germany}

% --- Abstract ---
\begin{abstract}
Food webs have been extensively studied from both ecological and mathematical aspects. However, most of the models studied in this area do not capture the effects of infectious diseases simultaneously. Recently, the idea of including an infectious disease in a food web model has been investigated. We study and simulate a small food chain consisting of only prey, predators, and apex predators governed by the generalized Lotka-Volterra equations, and we implement the Susceptible-Infected-Recovered (SIR) model on only one of the species at a time in the food chain. To study the effects of an infectious disease on the food chain, we introduce a new parameter that increases the predation rate by a factor of $w$ and decreases the hunting rate by a factor of $1/w$ for infected species. When the infectious disease is present in predators, we observe that predators do not become extinct under any set of parameters; however, an oscillation in their population size occurs under some circumstances, which we do not observe in ordinary SIR or the generalized Lotka-Volterra equations alone. When an infectious disease is present in apex predators, oscillations in the population size do not happen; but if the set of parameters is in a specific range the apex predators may become extinct. Furthermore, the chance of survival of the community, known as community persistence, increases for the predators and decreases for the apex predators.
\end{abstract}

\maketitle

\section{Introduction}
\label{intro}
%%%%%%%%%%%%%%%%%%%%%%%%%%%%%%%%%%%%

\vspace{0.5cm}

A food web is a fundamental concept in ecology that represents the intricate network of feeding relationships among different species within an ecosystem. It helps us understand how organisms interact with one another in terms of predation, competition, and energy transfer. Understanding food webs is crucial for several reasons. Firstly, it allows ecologists to gain insights into the stability and resilience of ecosystems, helping them predict how environmental changes or disturbances may impact species populations. Secondly, it aids in the conservation and management of biodiversity by highlighting the key species and their roles in maintaining ecological balance \cite{rooney2012integrating,dunne2006network}.

In recent ecological literature, there has been growing interest in studying food web structure and parasites together \cite{lafferty2008parasites}. Also, There has been an argument on how parasites should be included in food webs \cite{selakovic2014infectious,hatcher2011parasites}. Here, rather than explicitly incorporating parasites as nodes in the food web, we consider them through the infectious diseases they cause in a focal host species. In this framework, parasites influence the food web indirectly by altering the interactions of infected individuals across trophic levels.

Epidemiological studies on food webs are of significant importance because they provide insights into the spread of infectious diseases in an ecosystem \cite{johnson2015infectious}. When diseases affect species in different environments, their dynamics can interact in complex ways. Studying epidemics in food webs allows us not only to understand how diseases can spread in a certain species, but also how it affects other species as well. It enables us to study the overall stability and functioning of ecosystems. It sheds light on the intricate relationships between disease dynamics and food web dynamics, which can have important implications for conservation and ecosystem management \cite{collinge2006disease, smith2009role}. Though many papers might consider spatial effects in food webs, here, we neglect the possible geographical stretch of our food web to avoid adding any more complexity to our problem.

In this study, we develop a compact three-layer food chain model to investigate how pathogen transmission within a single focal species propagates and influences the dynamics of the entire food chain. Unlike previous studies, which have primarily examined pathogen effects when the infected species occupies either the basal or top trophic level \cite{de2015food, mbava2017prey}, our framework provides a systematic approach to study the infected species' and their population dynamics in the food chain. We introduce a new parameter, $w$, which quantifies the ecological impact of infected individuals on trophic interactions. This parameter provides a unified way to characterize pathogen-induced changes in species interactions and allows us to systematically compare different infection scenarios within the same modeling framework. Using this approach, we examine how disease transmission in predators or apex predators propagate and alter population dynamics, affects species persistence and extinction risk. Understanding these cascading effects is essential for assessing ecosystem stability and biodiversity, particularly because pathogen-driven extinctions may contribute to species vulnerability and conservation status \cite{mace2008quantification}.

When the infectious disease affects the predator population, predators persist across the entire parameter space explored and do not undergo extinction. Instead, for a subset of parameter values, the system exhibits sustained oscillations in predator abundance—a dynamical behavior that does not arise in either the classical SIR model or the generalized Lotka--Volterra equations when considered independently. In contrast, when the disease infects the apex predator, no oscillatory dynamics are observed. However, within a specific parameter regime, the apex predator may become extinct. Moreover, the overall persistence of the community increases when predators are infected but decreases when the apex predators are the infected trophic level. Our results not only agree with previous findings in limiting cases \cite{pedersen2007infectious, mougi2022infected}  but also extend them by providing a general framework for analyzing pathogen-mediated effects throughout a food chain.

\section{Our Model}

\begin{figure}
    \centering
    \includegraphics[width=\linewidth]{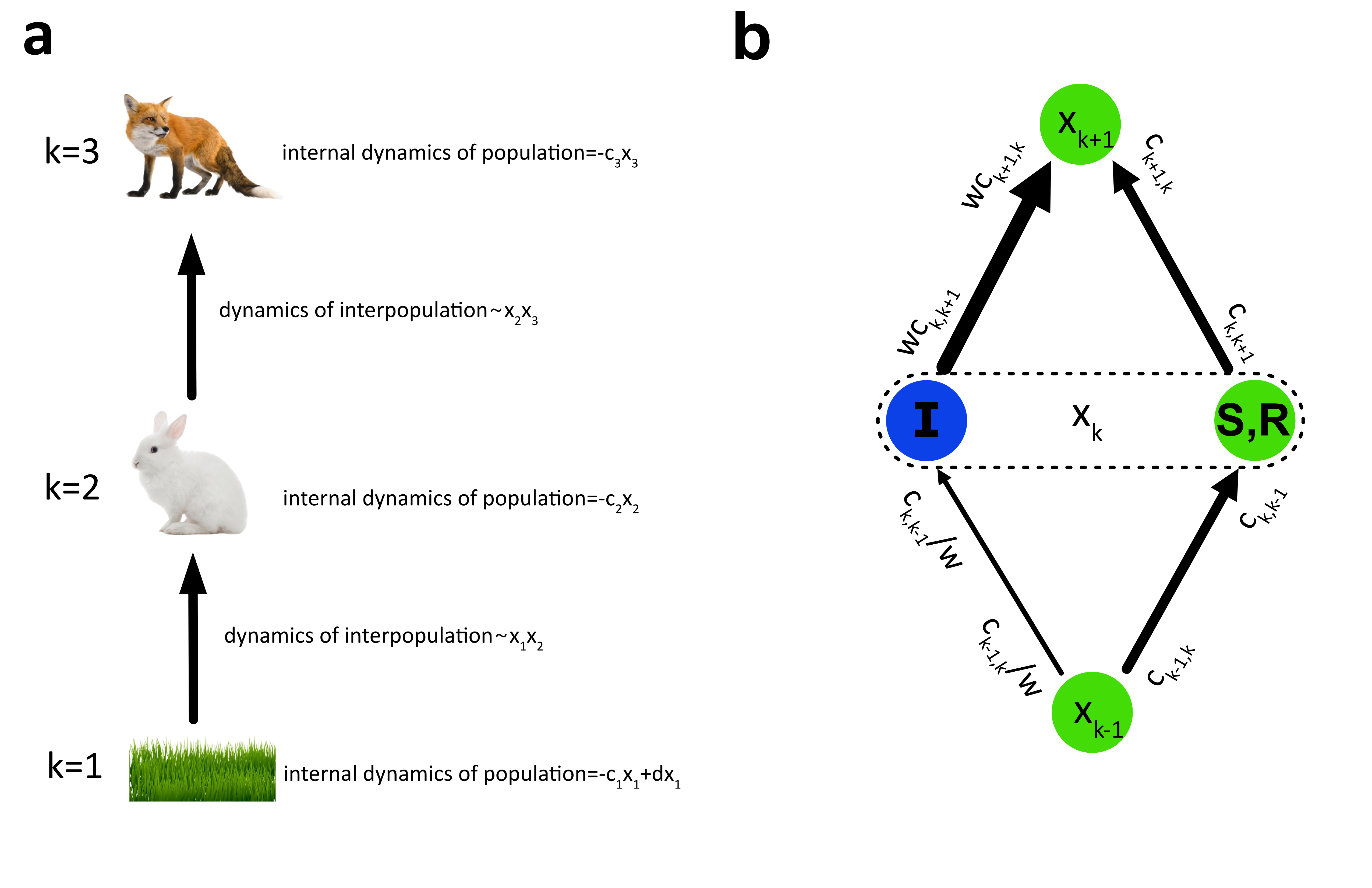}
    \caption{\textbf{A schematic figure of the model}. (\textbf{a}) indicates only the population dynamics in a food chain while (\textbf{b}) shows how the interpopulation dynamics change due to the presence of an infectious disease in layer k and the thickness of the arrows shows the strength of the interaction.}
    \label{fig:schematic}
\end{figure}

Here, we introduce our model, which couples two interacting dynamical processes that evolve simultaneously: food chain dynamics and infectious disease dynamics. These processes influence one another through the ecological interactions of infected individuals. Specifically, the model consists of

\begin{itemize}
\item \textbf{Food-chain dynamics:} a generalized Lotka--Volterra model \cite{diehl2000effects}.
\item \textbf{Infectious disease dynamics:} a Susceptible--Infected--Recovered (SIR) model \cite{kermack1927contribution}.
\end{itemize}

A variety of mathematical models have been proposed to describe food chain dynamics \cite{gross2009local}. In this work, we adopt the generalized Lotka--Volterra model \cite{diehl2000effects} as the underlying description of trophic interactions. This model captures the essential features of a three-species food chain while remaining sufficiently simple to allow for a comprehensive dynamical analysis, including a well-characterized phase diagram \cite{hsu2015analysis}. Since our primary objective is to investigate how infectious diseases alter food chain dynamics, employing a minimal yet representative ecological model enables us to isolate and quantify the effects of pathogen transmission without introducing unnecessary ecological complexity.

Throughout this paper except for the community persistence, we restrict our analysis to parameter regimes for which the disease-free food chain admits a stable, nontrivial equilibrium, i.e., all three species coexist. Under these assumptions, the food chain dynamics in the case of a disease-free food chain are governed by the following system of equations.

\begin{equation}\label{GLV}
    \begin{split}
        \dot{x_1}=& dx_1 -c_{12} x_1x_2 - c_1x_1^2 \\
 \dot{x_2}=& c_{21} x_1x_2 - c_{23}x_2x_3-c_2 x_2 \\
\dot{x_3}=& c_{32}x_2x_3  - c_3x_3  
    \end{split}
\end{equation}

where $c_{ij}$ is a constant that determines the strength of the interaction between the trophic levels in the food chain (see Fig.~\ref{fig:schematic}). The functions $f_i$ are subsequently modified to incorporate the effects of infectious disease while preserving the general structure of Eq.~(\ref{GLV}).

To model disease transmission, we partition the population of a single species into three epidemiological compartments: susceptible ($x_{k,S}$), infected ($x_{k,I}$), and recovered ($x_{k,R}$), where $k$ denotes the $k$th trophic level and such that $x_{k,S}+x_{k,I}+x_{k,R}=x_k.$
The disease dynamics are described by the classical SIR model:

\begin{equation}\label{SIR} 
\begin{aligned} 
\dot{x}_{k,S}=&-\frac{\beta x_{k,S}x_{k,I}}{x_k} \\ \dot{x}_{k,I}=&\frac{\beta x_{k,S}x_{k,I}}{x_k} -\gamma x_{k,I} \\ 
\dot{x}_{k,R}=&\gamma x_{k,I}, \end{aligned} 
\end{equation}

The SIR equations describe only the epidemiological dynamics within the selected species and do not account for trophic interactions. The coupling between epidemic and food chain dynamics is introduced in the next subsection by modifying the generalized Lotka--Volterra equations. Throughout this work, we assume that the pathogen infects only a single trophic level at a time, namely either the intermediate predator or the apex predator.

Simply combining the generalized Lotka--Volterra and SIR models does not produce any meaningful coupling between ecological and epidemiological dynamics. In such a formulation, the disease merely partitions the population into susceptible, infected, and recovered compartments, while trophic interactions remain unchanged. Consequently, the epidemic has no direct ecological effect beyond redistributing individuals among these compartments.

To establish a genuine coupling between the two dynamics, we distinguish infected individuals from susceptible and recovered individuals through their trophic interactions. Experimental and theoretical studies have shown that infected prey are often weaker, slower, and more susceptible to predation than healthy individuals \cite{lopez2023healthy,hatcher2011parasites,smith2009role,von2023computational}. Motivated by these observations, we introduce a new parameter, $w$, which quantifies the effect of infection on interaction strengths.

Throughout this work, we assume that susceptible and recovered individuals interact identically with the rest of the food chain. In contrast, infected individuals experience modified interaction coefficients. Specifically, predators consume infected individuals at an enhanced rate,
$
c_{k,k+1}(I)=wc_{k,k+1}(S)=wc_{k,k+1}(R), 
$  
while infected individuals hunt species in the lower trophic level less efficiently,
$
c_{k,k-1}(I)=\frac{1}{w}c_{k,k-1}(S)=\frac{1}{w}c_{k,k-1}(R).
$
Here, the superscript denotes the epidemiological compartment associated with the interaction coefficient (see Fig.~\ref{fig:schematic}). Thus, $w>1$ represents an increased vulnerability of infected individuals to predation together with a reduced hunting capacity, while $w=1$ recovers the standard generalized Lotka--Volterra model.

The coupled system consisting of Eqs.~(\ref{GLV}) and (\ref{SIR}) defines the complete model studied in this paper; Table \ref{tab:parameters} summarizes the variables and parameters of our model. 

For illustration, when the pathogen infects the second trophic level (predator), the coupled food chain dynamics become:

\begin{equation}\label{mixed} 
\begin{aligned}  \dot{x_1} =& dx_1 -\sum\limits_{i=S,I,R} [c_{12}(i) x_1x_{2i}] - c_1x_1^2 \\ 
 \dot{x_2} =& \sum\limits_{i=S,I,R}[ c_{21}(i) x_1x_{2i} - c_{23}(i)x_{2i}x_3-c_2 x_{2i}]\\ 
 \dot{x_3} =& \sum\limits_{i= S,I,R} [c_{32}(i)x_{2i}x_3] - c_3x_{3} \\ 
\end{aligned} 
\end{equation}

More information about this set of Eqs. (\ref{mixed}) is discussed in Section \ref{results} and the results are shown in Figures \ref{fig:subpopulationstime} and \ref{Fig:oscil}.

When the pathogen instead infects the third trophic level (the apex predator), the new food chain dynamics follow the following set of equations:

\begin{equation}\label{mixed_layer3}
\begin{aligned}
 \dot{x_1} =& dx_1 - c_{12}x_1x_2 - c_1x_1^2 \\
 \dot{x_2} =& c_{21}x_1x_2 - c_2x_2 - \sum\limits_{i=S,I,R} [c_{23}(i)x_2x_{3i}] \\
 \dot{x_3} =& \sum\limits_{i= S,I,R} [c_{32}(i)x_2x_{3i} - c_3x_{3i}] \\
\end{aligned}
\end{equation}

It should be noted that although the logic behind the modified equations  \ref{mixed} and \ref{mixed_layer3} is the same, the results obtained are rather different. We also discuss this in Section \ref{results} and the results are shown in Figure \ref{fig:extinct}.

\begin{table}[htbp]
\centering
\small
\begin{tabular}{ l p{5cm} } % Replace X with a fixed-width paragraph column
\hline\hline
\textbf{Parameter/Variable} & \textbf{Description} \\
\hline
$c_{i,i-1}(k), c_{i,i+1}(k)$ & Hunting rate and predation rate of the $k$-th subpopulation of species $i$. \\
\hline
$c_{i}, d$ & $c_i$ is mortality rate of species $i$. $d$ is the linear growth rate for species 1. \\
\hline
$\beta, \gamma$ & The rate of infection and recovery of the disease in species 2 or 3. \\
\hline
$w$ & The infection parameter on the infected species. \\
\hline
$x_{iS}, x_{iI}, x_{iR}$, $x_i$ & The susceptible ($x_{iS}$), infected ($x_{iI}$), recovered ($x_{iR}$), and the total population ($x_i$) in the food chain ($i$-th species). \\
\hline\hline
\end{tabular}
\caption{Variables and Parameters of the model.}
\label{tab:parameters}
\end{table}

\section{Results}\label{results}
To illustrate the dynamical behavior of the model, we numerically integrated the coupled equations using the fourth-order Runge-Kutta (RK4) method, for the following baseline parameter values: $d=0.12$, $c_{2}=0.2$, $c_{3}=0.27$, $c_1=0.002$, $c_{12}=0.002$, $c_{21}=0.003$, $c_{32}=0.004$, and $c_{23}=0.006$. For these parameter values, the asymptotic disease-free dynamics is independent of the initial conditions as well as other parameter values \cite{hsu2015analysis}. Unless otherwise stated, which is the case in community persistence, the same parameter set is used throughout Figures 2, 3 and 4.

 \subsection{Steady state}
Here, we study the steady state of the ecosystem when the predator is infected, Eqs. (\ref{mixed}). To determine the steady-state population (if one exists), we set $\dot{S}=0$, $\dot{I}=0$, and $\dot{R}=0$.
Thus, the following equalities hold at equilibrium.

\begin{equation}
\begin{aligned}
\dot{x}_{2S} &= 0 \longrightarrow c_{21}\left(1+\frac{x^*_{2R}}{x^*_{2S}}\right) x_{1} - c_{23}x_{3} - c_2 \\
&\qquad - \frac{\beta x^*_{2I}}{x^*_{2S}+x^*_{2I}+x^*_{2R}} = 0,\\
\dot{x}_{2I} &= 0 \longrightarrow \frac{c_{21}x_{1}}{w} - wc_{23}x_{3_ - c_2} \\
&\qquad + \frac{\beta x^*_{2S} }{x^*_{2S}+x^*_{2I}+x^*_{2R}} - \gamma = 0,\\
\dot{x}_{2R} &= 0 \longrightarrow -c_{23}x_{3} - c_2 + \frac{\gamma x^*_{2I}}{x^*_{2R}} = 0,
\end{aligned}
\end{equation}

where  $x^*_{2S}$, $x^*_{2I}$, and $x^*_{2R}$ are the susceptible, infected, and recovered populations at infectious disease equilibrium. Clearly, $x_{2}=x^*_{2S}+x^*_{2I}+x^*_{2R}$. If $\frac{x_{2I}}{x_{2R}}=const$, from the third equation we conclude that $x_3$ is constant, and thus $x_2$ and $x_1$ are constant as well. Hence the system is at equilibrium. Conversely, if the system is at equilibrium, then $\frac{x_{2I}}{x_{2R}}=const\geq 0$.
It can be shown that if $\beta \geq \left(\sqrt{AK} + \sqrt{(1+K)(A - B - c_2)}\right)^2$, and $Y > AK$, then there is no fix point at all, as we show in white region of the panel \textbf{b} of Figure \ref{Fig:oscil}, where

\begin{equation}
\begin{aligned}
&A = c_{21}x_{1} \\
& B = c_{23}x_{3} \\
&K=\frac{\gamma}{B+c_2} \\
&Y = \gamma + c_2 + wB - \frac{A}{w}.\\
\end{aligned}
\end{equation}

Notice $x_1$ and $x_3$ are themselves a function of $\beta$, $\gamma$, and $w$, however bounded. Moreover, we stress that collective aspects of an ecosystem such as populations is unchanged for $w=1$, since the pathogen has no effect in the ecosystem dynamics.

\subsection{Predator}
Some experimental and theoretical findings show that an infectious disease can persist in a food chain \cite{anderson1986invasion}. The following Figure \ref{fig:subpopulationstime} sketches the parameter space for which a disease on a second species can linger in our 3-layer food chain (prey, predator, apex predator).  This is, of course, no surprise, since the birth of uninfected individuals contributes to the susceptible population.

\begin{figure*}[htb!]
    \centering
    \includegraphics[width=0.7\linewidth]{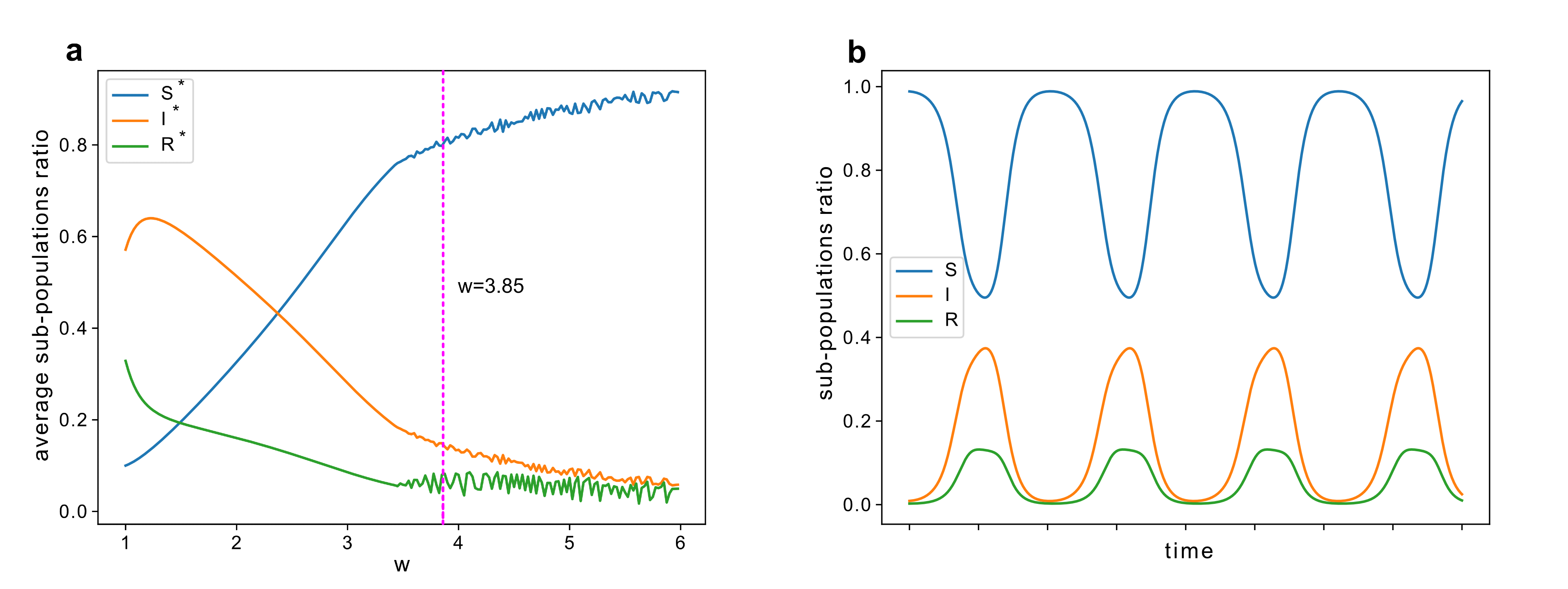}
    \caption{ \textbf{ The subpopulations of the predator when infected}. (\textbf{a}) shows the average ratio with respect to the overall species population of time-averaged susceptible ($S^*$) and infected  ($I^*$) and recovered  ($R^*$) subpopulations for different values of $w$ and $\frac{\beta}{\gamma}=10$ in the predators. The disease  begins to disappear as $w$ increases to a large enough value. (\textbf{b}) shows that the subpopulations in their equilibrium states oscillate for $w=3.85$ (the purple line in panel \textbf{a}) as a function of time.   }
    \label{fig:subpopulationstime}
\end{figure*}

\begin{figure*}[!htb]
    \includegraphics[width=0.7\linewidth]{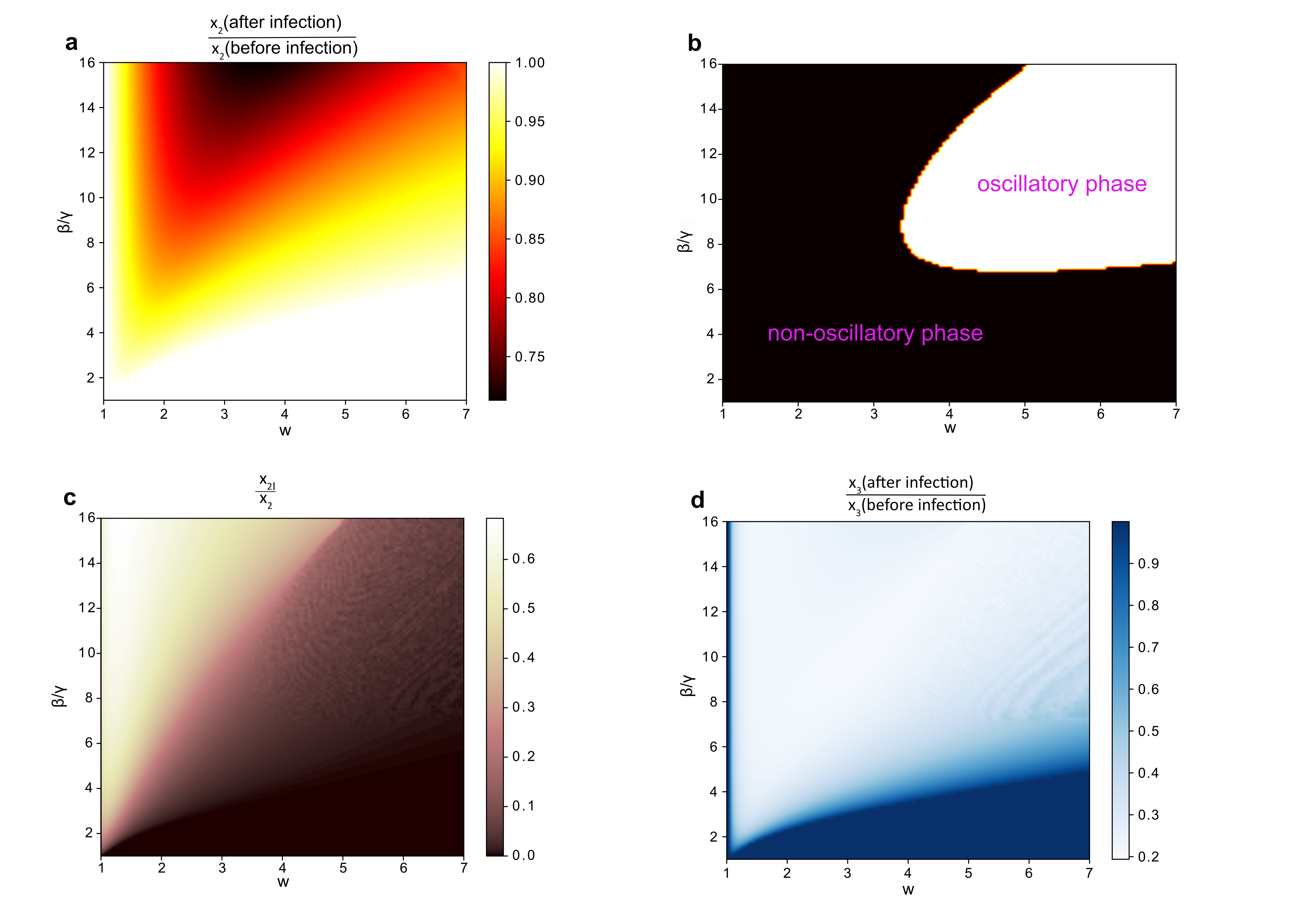}    
    \caption{ \textbf{General behavior of SIR on predator}. This panel depicts the influence of an infectious disease affecting the predator species on the population dynamics of the ecosystem, plotted against parameters $w$ and  $\frac{\beta}{\gamma}$. (\textbf{a}): The ratio of the predator population after the disease outbreak to its population before the outbreak. (\textbf{b}): Regions of oscillatory (white) and non-oscillatory (black) behavior in the predator population's final state. (\textbf{c}): The proportion of the predator population that is infected. (\textbf{d}): The ratio of the third species' population after the infection to its initial population, demonstrating the indirect impact of the disease.}  
    \label{Fig:oscil}
\end{figure*}

Clearly, two qualitatively different outcomes are possible: the infectious disease either dies out over time or persists in the food chain. We are primarily interested in the latter case. A natural question is: if the disease persists in the second species, to what extent does it alter the population relative to the disease-free equilibrium? Figure \ref{fig:subpopulationstime} sheds light on this matter.  The second population drops as expected; however the situation for the third species is not as clear as for the second, since the resources are reduced, and yet resource availability increases because of the infected population in the predators. This causes an increase in the third population. Panel \textbf{a} of Figure \ref{fig:subpopulationstime} shows that the second-species population undergoes a phase transition, which we investigate further in Figure \ref{Fig:oscil}. For an appropriate set of parameter values, the system exhibits oscillatory dynamics in the presence of the infectious disease as shown in Panel \textbf{b} of Figure \ref{fig:subpopulationstime} as well as Panel \textbf{b} of Figure \ref{Fig:oscil}. When the infected population is low, the third population is also low. Consequently, predation pressure from the third population is weak because $x_3$ is small. As a result, both interaction terms, $c_{32}x_3x_2$ and $c_{23}x_3x_2$, are small. This weak predation facilitates the spread of the disease, leading to an increase in the infected population and, consequently, in the second population, which is initially below its equilibrium level.

As the infected population grows, the third population also increases, as it is far from its equilibrium. This trend continues until the second population reaches a limit and begins to decline due to a reduced hunting rate (high $x_{2I}$) and increased predation from the now-larger third population (Now $x_3$ is overblown) until both decrease and reach the previous state.  

 Our simulations show that, when the second species is infected, there is no parameter regime in which the ecosystem collapses, i.e., the second species goes extinct.
  Considering that infectious diseases are rarely the main cause of extinction, this result is not illogical \cite{smith2006evidence}. However, for some values of $w$, $\beta$, and $\gamma$, the second population (and consequently the third one) starts to oscillate, as shown in Figure \ref{fig:subpopulationstime}, this is a new behavior different from the disease-free situation. Now, we investigate further and sketch the populations in Figure \ref{Fig:oscil}, for the parameter $w$, when the second species is infected by $\beta$, and recovered by $\gamma$ parameters.  

 It is evident from  panel \textbf{a} of Figure \ref{Fig:oscil} that the second species does not become extinct as a result of the introduction of an infectious disease. We can infer from panel \textbf{c} of figure \ref{Fig:oscil} that the pathogen cannot infect the entire population at the same time. Furthermore,  panel \textbf{d} of figure \ref{Fig:oscil} shows that despite the fact that the third species population does not reach zero, it is heavily affected by the disease. Also, the ratio is clearly 1 when the disease is not present, e.g. for the case of $w=1$.   So the third species may go through quasi-extinction, meaning that the population drops below a certain threshold \cite{ginzburg1982quasiextinction}. 
 Figure~\ref{Fig:oscil} also shows that the average quantities displayed in panels \textbf{a}, \textbf{c}, and \textbf{d} remain essentially unchanged as the system transitions from the black region to the white region, panel \textbf{b}.

\subsection{Apex Predator}
Likewise, we can ask the same question when the third species is infected. In this case, however, the outcome is different: for certain parameter regimes, the apex predators become extinct, as shown in the following Figure \ref{fig:extinct}, Panel \textbf{b}; While panels \textbf{a} and \textbf{c} show how the populations of the predator and apex predator are affected by pathogen infection of the apex predator. 

\begin{figure*}[htb!] 
    \centering
       \includegraphics[width=\linewidth]{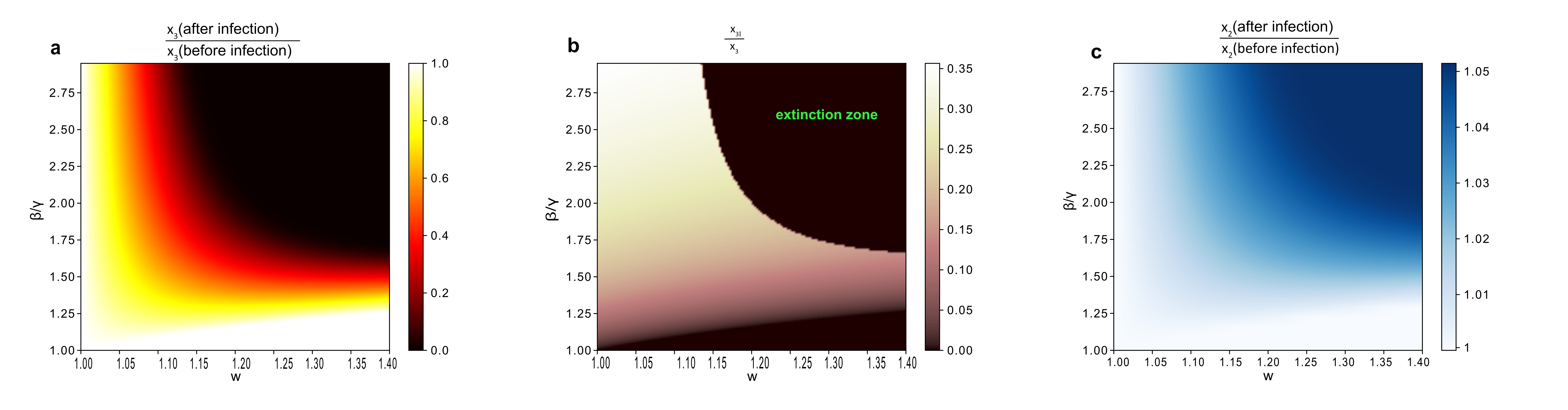}     
    \caption{ \textbf{General behavior of SIR on apex predator.} These panels depict the populations when the third species is suffering from an infectious disease as a function of $w$ and $\frac{\beta}{\gamma}$. (\textbf{a}) shows the population of apex predators during an outbreak within their own trophic level. The black area shows the extinction of the apex predators. (\textbf{b}) shows the ratio of infected population to the whole predator population. (\textbf{c}) shows the ratio of the predator species population to its original population before the infectious disease. This is  similar to the first two panels.}
    \label{fig:extinct}
\end{figure*}

Since apex predators might become extinct as a result of an infectious disease, there is a clear contrast between predators and apex predators when they suffer from an infectious disease. In addition, this confirms the finding in \cite{hollings2016disease} that apex predators are more at risk of disease-induced extinctions. As seen in panel \textbf{b} of Figure \ref{fig:extinct}, the extinction zone can be reached from below the extinction zone, where there is a relatively low infected population. Notice that, unlike the overall population, the infected population can go to zero for other reasons, such as a low infection rate in the species, as we can see in panel \textbf{b} of Figure \ref{fig:extinct}. In panel \textbf{c} of Figure \ref{fig:extinct}, we can see that the predator population does not change significantly. However, notice that our dynamics are closer to reality when species populations are large enough, since in small populations inbreeding depression \cite{saccheri1998inbreeding} and genetic drift \cite{masel2011genetic} will also play a role in population dynamics. 

\subsection{Community Persistence}
\begin{figure*}[htb!] 
    \centering
    \includegraphics[width=\linewidth]{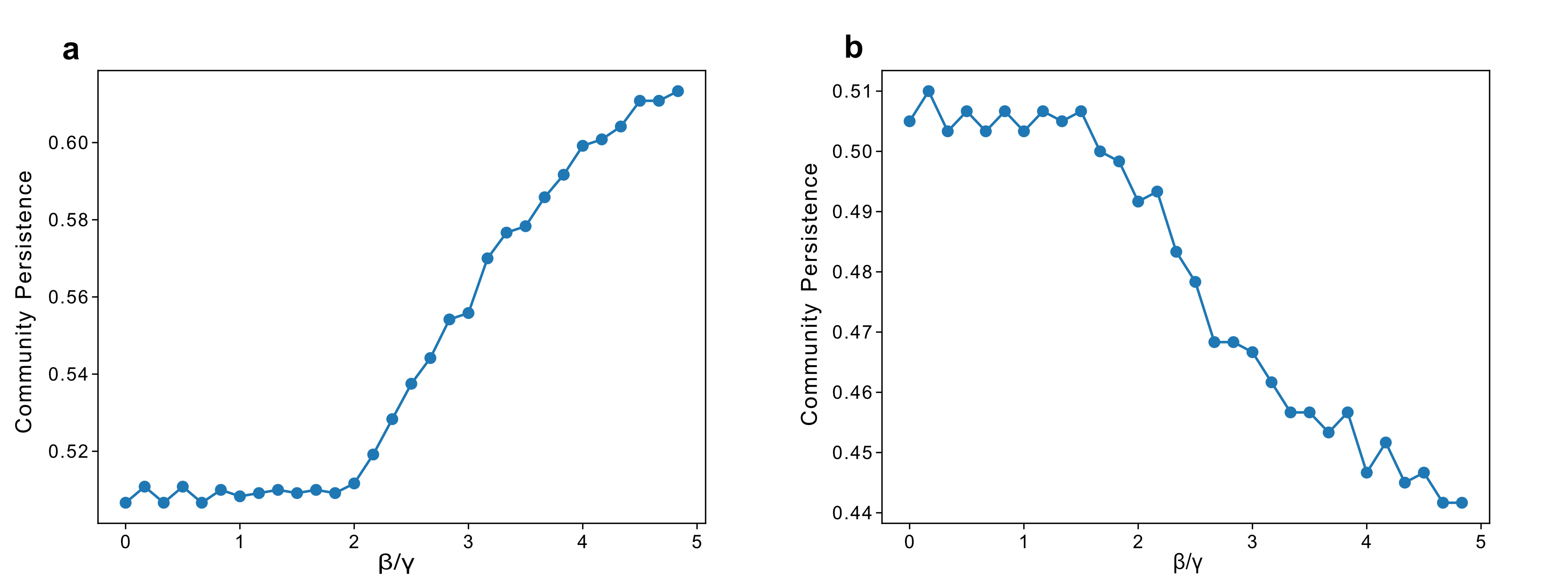}
    \caption{\textbf{Community persistence change due to the disease.} (\textbf{a}) Community persistence of the predator population  $(w=3)$. The community persistence increases with $\frac{\beta}{\gamma}$. (\textbf{b}) Third population community persistence ($w$=3). Community persistence decreases here as opposed to panel \textbf{a}, where community persistence increases.}
    \label{fig:communitypersistence}
\end{figure*}

Although the population dynamics are important, their exact solutions are hard to obtain and not always necessary. From an ecological perspective, evaluating food chain stability is often more realistic. One way to quantify this is known to be {\it community persistence}, which is the probability of not having any of the species extinct starting from a random point in the parameter space \cite{mougi2021diversity}. Here, this probability is estimated over an ensemble of 1,000 iterations, and parameters are sampled from a Gaussian distribution centered on the previously specified values, with a standard deviation set to one-half of the mean. We plot community persistence when the  predators and apex predators suffer from an infectious disease separately, see Figure \ref{fig:communitypersistence}. This can be used as a metric to show how stable our ecosystem is. As we have seen, the infectious disease on the predators will not wipe out the ecosystem. Thus, it is not unreasonable to expect that it might actually increase the ecosystem's stability. Indeed, as shown in panel \textbf{a} of Figure \ref{fig:communitypersistence}, this is the case, which might seem counterintuitive. However, another study involving more intermediate layers in a food chain also demonstrated that infectious diseases across various species lead to higher community persistence \cite{mougi2022infected}, which a with our panel \textbf{a} of Figure \ref{fig:communitypersistence}.

 Similarly, since for some sets of parameters the apex predators become extinct, we can expect that community persistence drops, and that is what we see in panel \textbf{b} of Figure \ref{fig:communitypersistence}. This shows that disease at the top level in food chains can decrease the stability of the food chain. This result confirms some experimental findings that show apex predators are more in danger of extinction from infectious diseases \cite{hollings2016disease} . So, we observe that the effects of an infectious disease at different trophic levels of food chains can have different effects on the food chain and its stability.

 If we consider community persistence to be an order parameter, then we see that a phase transition begins to occur as the disease starts to spread. This phase transition is different from the oscillation in Figure \ref{Fig:oscil} or the extinction of the third species in Figure \ref{fig:extinct}, and it is related to the existence of the disease.

%%%%%%%%%%%%%%%%%%%%%%%%%%%%%%%%%%%%
\section{Discussion and concluding remarks}
In this work, we proposed a minimal framework for coupling epidemic and ecological dynamics by integrating a classical SIR model with a three-layer generalized Lotka--Volterra food chain. Rather than introducing pathogens as explicit components of the food web, we modeled their ecological impact through changes in the trophic interactions of infected individuals. This coupling was achieved through a single parameter, $w$, which quantifies how infection modifies predation efficiency and vulnerability. The simplicity of this formulation makes it possible to isolate the ecological consequences of disease while retaining the essential features of both epidemic and food chain dynamics.

Our results demonstrate that the trophic position of the infected species plays a decisive role in determining the long-term behavior of the ecosystem. When the intermediate predator is infected, the coupled system exhibits sustained oscillations that do not arise in either the classical SIR model or the generalized Lotka--Volterra equations considered separately. Moreover, the persistence of the ecological community increases under this infection scenario. In contrast, infection of the apex predator does not generate oscillatory dynamics but can drive the apex predator to extinction for a range of parameter values, thereby reducing overall community persistence. These findings highlight that the ecological consequences of disease depend not only on pathogen characteristics but also on the position of the host species within the food chain.

The proposed framework is intentionally minimal and therefore has several limitations. We considered a simple three-species food chain with a single infected trophic level and classical SIR disease dynamics. In addition, the effects of infection were represented through a single coupling parameter that modifies trophic interaction strengths. Real ecosystems may include alternative prey, omnivory, adaptive foraging, seasonal variation, spatial heterogeneity, species movement, and more complex epidemiological processes such as latency, waning immunity, or multiple pathogens \cite{cai2015avalanche,chen2013outbreaks,zarei2019exact,ghanbarnejad2022emergence,hebert2025one} and host species. Incorporating these mechanisms would provide a more realistic description of disease-mediated ecological dynamics.

Despite these simplifications, the present model provides a systematic and flexible framework for investigating how infectious diseases influence food chains through trophic interactions. Because the coupling mechanism is independent of the specific ecological or epidemiological model, it can be readily extended to larger food webs, alternative disease dynamics, or more realistic ecological interactions. We hope that this work motivates further studies on the interplay between epidemic processes and ecological networks, ultimately contributing to a better understanding of pathogen-driven changes in ecosystem stability and biodiversity.
%%%%%%%%%%%%%%%%%%%%%%%%%%%%%%%%%%%%

\medskip
\bibliographystyle{IEEEtran}
\bibliography{biblo}
\end{document}